# Field induced reversible control of visible luminescence in ZnO nanostructures

Manoranjan Ghosh[1] and A.K.Raychaudhuri[2]
*DST Unit for Nanoscience, S.N.Bose National Centre for Basic Sciences,*
*Block-JD, Sector-3, Salt Lake, Kolkata-700 098, INDIA*

[1]Email: mghosh@bose.res.in
[2]Email: arup@bose.res.in

**Abstract:** In this work, a reversible control over the visible luminescence of phosphor ZnO is achieved by the application of a few volts (< 5V) to nanostructured ZnO film sandwiched between ITO and $LiClO_4$/PEO solid electrolyte. Since ZnO is a good absorbing material but ITO-Glass substrate is transparent in the range of 320–375 nm, the ZnO-ITO interface has been illuminated by a 345 nm light passing through the ITO glass substrate and the emitted light near 545 nm has been collected in reflection mode. When the solid polymer electrolyte is negatively biased, up to 107 percent enhancement in the visible luminescence has been observed, whereas a complete quenching (87 percent reduction) of the visible emission has been seen when it is positively biased. The sharp NBE emission (near band edge) in the ultraviolet region however has no dependence within the voltage range we are working. The luminescence output follows variations in the applied voltage up to a maximum frequency limited by the response time of the device which lies within the range of 20-30 sec. The upward band bending created by application of negative bias to the conductive polymer populates increasing number of defect levels above the Fermi level which enhances the visible photoluminescence and vice versa.

**Introduction:** Over the past decade, tremendous progress in research has been achieved for the use of ZnO as an optoelectronic material. ZnO, a wide band gap (3.3 eV) semiconductor with wurtzite structure (P63mc) is a well-known phosphor for emission in the visible [1-4]. In particular, nanostructured ZnO that can be fabricated at temperatures less than $100^0$C can be used for innovative optoelectronic devices. The room temperature photoluminescence (PL) of ZnO nanostructures (size ~ 10-15 nm) typically exhibits a sharp emission in the near UV range (originating from excitonic mechanism) and a broad emission band in the visible region of the spectrum which is linked to surface defects [5]. Control of this visible emission is technologically as well as academically very relevant. In a earlier publication from our group, we have established that the visible emission depends on many factors like size and shape of the material, surrounding media and type of defects involved in the emission processes [6]. The influence of these factors on this broad visible luminescence is reported and the positively charged oxygen vacancies were claimed to be responsible for this emission in many circumstances [7,8]. Fine control of this visible emission by the application of a voltage is a very important step for its application. Recently it is demonstrated that the charge environment surrounding the nanoparticles is capable of controlling this visible emission [9]. Quenching of fluorescence from the phosphors like ZnS by application of few volts has been reported [10]. In a recent report quenching of the visible emission and enhancement of the UV emission by application of an electric field has been observed [11]. To our knowledge, a reversible control (quenching as well as enhancement) of the visible luminescence of ZnO nanostructures by application of very small voltage (< 5 V) has not been achieved earlier. Below we describe the phenomenon of Electric Field Controlled Photoluminescnce or Electrophotoluminescence on a device based on nanostructured ZnO.

**Device fabrication and the arrangement:** The phenomenon of field control of visible luminescence is investigated on a nanostructured ZnO film. In the following we give description of the device consisting of ITO/ZnO nanostructures/polymeric electrolyte, in which we carried out our investigation. Figure 1 shows the experimental arrangement adopted for the fabricated device. ZnO nanoparticles of size ~ 10 -15 nm have been synthesized by acetate route described in ref [12, 13]. The nanostructured ZnO film (thickness varying from 200-300 nm) is obtained by adding these nanoparticles dispersion in ethanol drop wise on a pre-decided area of a commercially available conducting ITO layer supported by a glass substrate. The ZnO dispersion has been spread over the surface uniformly and dried at room temperature. Equal numbers of drops were added to obtain films of similar thickness repeatedly. To assure the presence of a ZnO layer all over the ITO surface, a thick film is deposited. After the ZnO film dries off, a gel containing 10:1 weight ratio of PEO (poly ethylene oxide, MW 1000) and *LiClO*$_4$ is deposited on the ZnO layer at room temperature. After 3-4 hours duration, the gel becomes solid and holds the electrical connections (through Cu wires) tightly as shown in figure 1.1. The ZnO-ITO interface has been illuminated by a 345 nm light passing through the ITO glass substrate and the emitted light (of wavelength 545 nm) has been collected in reflection mode. The biasing arrangement of the device is shown in figure 1. The device is said to be positively (negatively) biased when the positive (negative) terminal of the source is connected to the *PEO/LiClO*$_4$ solid electrolyte respectively through a copper wire.

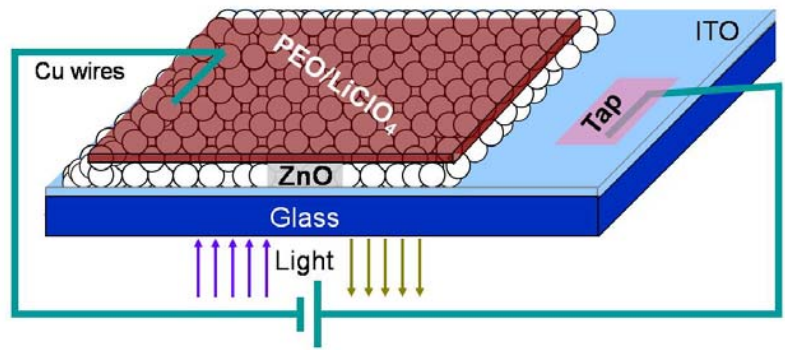

Figure 1: Illustration of the device fabricated. The biasing arrangement demonstrated here leads to enhancement of intensity and vice versa.

UV light is allowed to pass through the ITO-Glass substrate from a Xenon arc lamp and the monochromator of a Spectrofluorimeter (Jobin Yvon Fluromax 3). The bias voltages were supplied by a Stanford Research System DS345 synthesized function generator and the device current is recorded by a Keithley, 6514 System Electrometer. The sizes of the nanoparticles were determined by JEOL High Resolution Transmission Electron Microscope (HR-TEM) operated at 200 KeV. TEM image of the collection of ZnO nanoparticles [figure 2 (a)] shows that the average size of the nanoparticles is ~10 nm. When the dispersion of such nanoparticles is deposited on a substrate agglomeration takes place. The average size of these agglomerated nanoparticles is found to be 40-50 nm [figure 2(b)] when imaged by Field Emission Gun Scanning Electron Microscope (FEG SEM). The thickness of such a film of ZnO nanoparticles is determined from the AFM image operated in contact mode with a scratch on the film as shown in figure 2(c). The average thickness measured at the step of ZnO-ITO was found to vary from 200 to 300 nm.

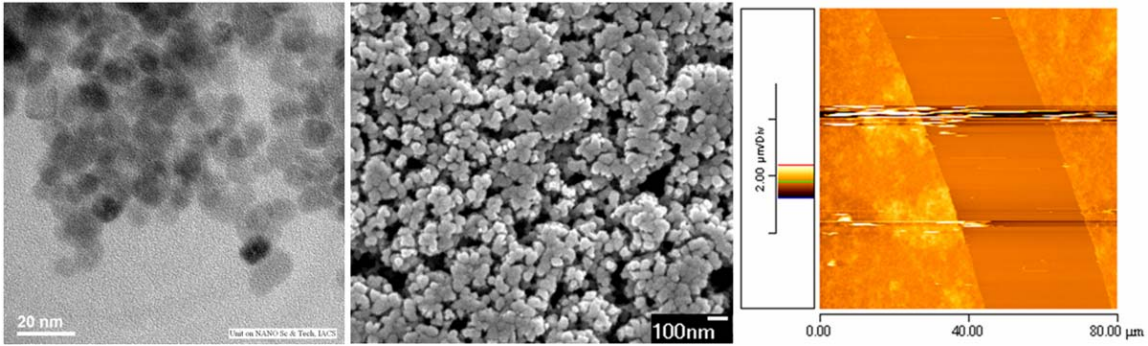

Figure 2: (a) TEM image of the ZnO nanoparticles on a grid. (b) SEM image of the nanostructured film obtained by depositing ZnO nanoparticles shown in (a) on a ITO substrate. (c) AFM image of the scratched film of ZnO nanostructures [shown in (b)] used for obtaining the film thickness. Thickness of the ZnO film investigated in this work varies from 200-300 nm as determined at the step of substrate-ZnO film.

**The excitation wavelength:** Now we first work out the suitable range of optical wavelength over which the device can perform. We find that 345 nm is the most suitable window for excitation of this device. In the range of 320 – 375 nm ZnO is a good absorbing material but ITO-Glass substrate is transparent as shown in figure 3. Therefore, UV light of wavelength in the range 320 – 375 nm can easily pass through the ITO glass substrate but will be absorbed fully by the thick ZnO layer. Simultaneously, the visible light emitted from ZnO can pass through the ITO glass substrate without much absorption. [shown in figure 4(a)], matches qualitatively with the data obtained for the colloidal ZnO nanoparticles. Therefore, one can excite the device through ITO glass face and the emitted signal can be collected in the reflection mode. This is an important observation for fabrication of any optoelectronic devices which is a combination of conductive ITO and a semiconductor ZnO, specially, for the kind of device which uses non-transparent top layer. More importantly we have achieved an undistorted visible PL signal by eliminating effect of the polymer electrolyte which gives a broad PL peak in the blue region. For that purpose we have used a thick ZnO film which completely absorbs the excitation light.

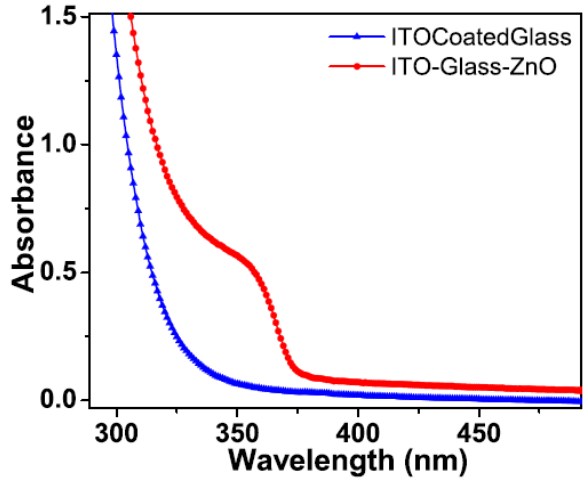

Figure 3: Absorption spectra of a bare ITO coated glass substrate and the other drop coated with a ZnO nanostructured film.

**Control of visible luminescence by application of voltage:** The main result of this work is the control of the photo-luminescence properties of the device fabricated in the way described in figure 1. In figure 4(a) we show the photoluminescence spectra of the device after the excitation by a light of wavelength 345 nm through the ITO layer for which the emitted photons were collected in reflection mode. Along with the UV peak at 400 nm, there is a broad peak at 545 nm. The photoluminescence spectra from the device

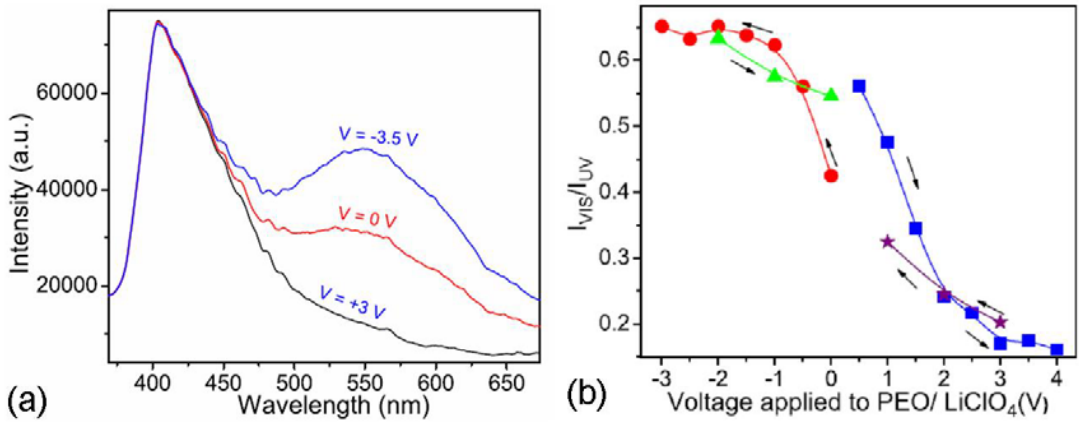

Figure 4: (a) Photoluminescence spectra showing reversible control of the visible luminescence near 544 nm by the application of voltage as indicated on the graph. The NBE emission around 400 nm remains unchanged by the application of voltage. (b) The intensity ratio of the visible (~400 nm) to the NBE emission (~544 nm) has been plotted with decreasing or increasing applied voltage as indicated on the graph. Reference of bias is on ITO layer.

But we observe a reduction in relative intensity of the surface related visible emission from ZnO due to surface passivation when kept sandwiched between the ITO and *PEO/LiClO$_4$* layer. The intensity ratio of the UV to visible emission of the ZnO nanoparticles in the device is 5 times lower than that of the colloidal ZnO nanoparticles. Now we apply a DC bias on the two terminals (ITO and *PEO/LiClO$_4$*) as shown in figure 1. When the illuminated surface (ITO-ZnO interface) is positive (biasing arrangement shown in figure 1), nearly 100 percent enhancement in the visible luminescence has been observed [figure 4(a)]. A complete quenching of the visible luminescence is possible by the application of negative bias at the illuminated surface. The intensity ratio of the visible (~400 nm) to the NBE emission (~544 nm) has been plotted with decreasing or increasing applied voltage as indicated on the graph [figure 4(b)]. There exists a hysteresis loop due to the delay in the voltage controlled visible luminescence process. In the current situation, time gap between two zero bias measurement was 20 minutes which is not sufficient for the complete relaxation of the excited state electrons.

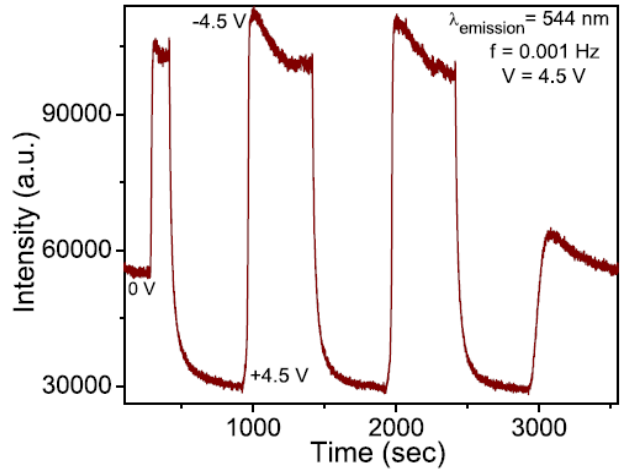

Figure 5: AC voltage response of the visible luminescence intensity as a function of time collected at 544 nm exactly follows the amplitude and the polarity of the applied voltage. Nearly 100% modulation of the visible luminescence is achieved.

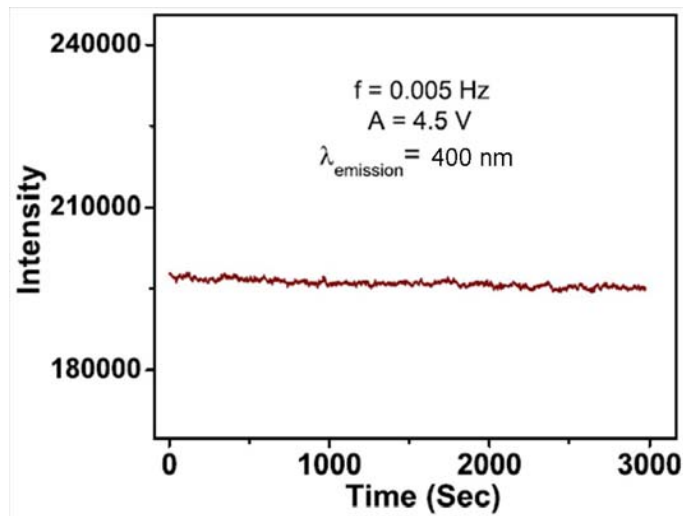

Figure 6: The NBE emission at 400 nm does not have any dependence on the polarity and amplitude of the AC voltage (shown in the same intensity range as in the case of visible emission).

The quenching and enhancement of the visible luminescence can be very nicely demonstrated if we apply an AC voltage of frequency 0.1 Hz and amplitude 4 V. For positive half cycles, up to 107 percent enhancement in the visible luminescence has been observed [figure 5] whereas a complete quenching (87 percent reduction) of the visible emission has been seen when the illuminated surface is negatively biased. The sharp NBE emission (near band edge) in the ultraviolet region however has no dependence on the applied voltage (figure 6). The response time of the luminescence output to the applied voltage lies within the range of 20 -30 sec. AC response of the luminescence output remains constant with a square wave of time period faster than 20 sec.

## Discussion:

The observation reported here essentially arises from the ZnO-electrolyte interface. To check that these effects are predominantly related to the semiconductor-electrolyte interface, we fabricated a device consisting of top Al layer in place of polymeric electrolyte (figure 8). There is a small decrease of visible PL at a much higher voltage when a positive bias is applied to the Al (ZnO is negatively biased). No enhancement is seen as in the case of electrolyte based junction.

The polymer-electrolyte junction upon application of an appropriate potential between the electrodes can change the oxidation/reduction condition. The negatively biased polymer electrode is reduced (electron addition) while the other oxidizes (electron removal). In forward bias the current is expected to be carried by the majority carrier (in this case, electrons, ZnO being n-type). In reverse bias (polymer-electrolyte negative, ZnO positive) two situations can arise. One can get electrons crossing into the ZnO from the electrolyte (deposited by anions, $ClO^{-4}$, which collect around the interface since the counter ion $Li^+$ is pulled to the negatively charged electrode). In addition, in certain range of bias (moderate reverse bias) another process can occur where a majority carrier from ZnO (electron) can cross over to the electrolyte side.

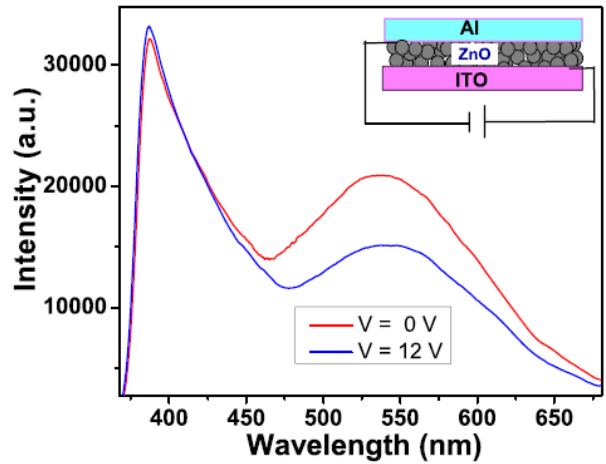

Figure 8: The PL spectra of a device with a conducting Al layer (replacing the *PEO/LiClO$_4$* layer) on top of ZnO film as shown in the graph. Reference of bias is on ITO layer.

When a reduced electrolyte state is at the same energy as the empty states in the semiconductor, i.e., when the states overlap, an electron can transfer to the semiconductor. An electron can transfer from the semiconductor to the electrolyte when an oxidized electrolyte state overlaps the filled states of the semiconductor. This overlap approach has been useful for qualitative discussion of the many combinations of semiconductors and redox couples that have been investigated experimentally.

It has been established that the visible emission from ZnO in the blue-green region originates from defects (oxygen vacancy in particular) which are located predominantly near the surface of the nanoparticles [6]. It is also seen that this emission is a composite of two broad lines located approximately at 2.2 eV (550nm) and at 2.5 eV (500nm). Emission band appearing around 550 nm (*P*2) has been suggested to originate from doubly charged oxygen vacancy ($V^{++}_o$) whereas singly charged oxygen vacancy ($V^{+}_o$) is responsible for the emission band around 500 nm (*P*1). The relative intensities of the two lines will depend on their relative occupancy and determines the exact position and overall intensity of the visible luminescence [6].

In a previous report from our group, we have shown how the defect related visible PL can be controlled by the control of the surface charge which in turn controls the band-bending at the surface of the ZnO nanoparticles [9]. We like to propose an explanation in line with the physical picture proposed earlier. We propose that when negative voltage applied to the electrolyte the $Li^+$ ions move away from the ZnO-electrolyte interface. This leads to formation of a negatively charged layer on the surface. This induces more positive charge on the ZnO surface leading to more band bending and thus strengthening the emission from the P2 band [9]. When a positive voltage is applied to the electrolyte the ZnO-electrolyte interface accumulates the $Li^+$ ions which induces more negative charge (or reduces the positive charge) in the surface region of ZnO and hence decreases the effect of band bending. This leads to reduction of the PL. We note that the enhancement of the PL occurs in the reverse bias condition when the ZnO is positively biased. This as stated earlier increases the depletion width. In this situation the junction acts like a capacitor. Since there is also a current flowing in the junction, the occupation of the defect sites may be affected. However, it appears that the primary effect of the bias is to redistribute the charges at the junction through the electrolytes.

**Conclusions:**
We have achieved a reversible control over the visible photoluminescence in ZnO nanostructures with the application of electric field through a solid polymer electrolyte. The sharp NBE emission (near band edge) in the ultraviolet region however has no dependence within the voltage range we are working. An electric double layer forms at the positively charged ZnO and polymer electrolyte interface. Band bending occurs in both directions (upward as well downward) when alternating voltage is applied. Relative occupancy of the defect levels is affected by this band bending which in turn controls the overall intensity of the visible luminescence. Although slow diffusion of ions for the formation of double layers may limit its application at high frequency, the outstanding dc performance and the easy control of the luminescence intensity would make the device attractive for future applications.

**Acknowledgement:** The authors would like to thank IACS for providing TEM facility and the Department of Science and Technology, India for the financial support as a unit for Nanoscience.